**Solid phase epitaxial growth of the correlated-electron transparent conducting oxide SrVO$_3$**

Samuel D. Marks,[1,*] Lin Lin,[1] Peng Zuo,[1] Patrick J. Strohbeen,[1] Ryan Jacobs,[1] Dongxue Du,[1] Jason R. Waldvogel,[1] Rui Liu,[1] Donald E. Savage,[1] John H. Booske,[2] Jason K. Kawasaki,[1] Susan E. Babcock,[1] Dane Morgan,[1] and Paul G. Evans[1]

[1] *Department of Materials Science and Engineering, University of Wisconsin, Madison, WI 53706, USA*

[2] *Department of Electrical and Computer Engineering, University of Wisconsin, Madison, WI 53706, USA*

*\* sdmarks2@wisc.edu*

SrVO$_3$ thin films with a high figure of merit for applications as transparent conductors were crystallized from amorphous layers using solid phase epitaxy (SPE). Epitaxial SrVO$_3$ films crystallized on SrTiO$_3$ using SPE exhibit room temperature resistivities as low as $5.2 \times 10^{-5}$ $\Omega$ cm and $2.5 \times 10^{-5}$ $\Omega$ cm, residual resistivity ratios of 2.0 and 3.8, and visible light transmission maxima of 0.89 and 0.52 for film thicknesses of 16 nm and 60 nm, respectively. SrVO$_3$ layers were deposited at room temperature using radio-frequency sputtering in an amorphous form and subsequently crystallized by heating in controlled gas environment. The lattice parameters and mosaic angular width of x-ray reflections from the crystallized films are consistent with partial relaxation of the strain resulting from the epitaxial mismatch between SrVO$_3$ and SrTiO$_3$. A reflection high-energy electron diffraction study of the kinetics of SPE indicates that crystallization occurs via the thermally activated propagation of the crystalline/amorphous interface, similar to SPE phenomena in other perovskite oxides. Thermodynamic calculations based on density functional theory predict the temperature and oxygen partial pressure conditions required to produce the SrVO$_3$ phase and are consistent with the experiments. The separate control of deposition and crystallization conditions in SPE presents new possibilities for the crystallization of transparent conductors in complex geometries and over large areas.

Keywords: epitaxial transparent conducting oxides, solid-phase epitaxy, strontium vanadate, phase selection in oxide synthesis

## I. INTRODUCTION

Transparent conducting oxides (TCOs) are key components of technologies requiring separate control of the electrical and optical properties of materials in devices including photovoltaic cells, displays, and smart windows.[1-4] These applications require low electrical resistivity and high optical transparency in the visible spectrum. Strontium vanadate ($SrVO_3$) is an emerging TCO material and has promising further applications in oxide electrodes, thermionic emission, and solid-oxide fuel cells.[5-10] Electronic-correlation-driven effective mass enhancements of the V $3d$ states push the plasma frequency of $SrVO_3$ into the ultraviolet, yielding metal conductivity and transparency to visible light.[8]

Crystalline $SrVO_3$ has the cubic perovskite structure in which V is at the center of O octahedra.[11,12] In addition to $SrVO_3$, the Sr-V-O phase diagram includes compounds with different Sr:V ratios and V oxidation states.[13] Deviations from Sr:V 1:1 stoichiometry, either at the nanoscale or overall, can lead to the formation of competing phases, including $Sr_3V_2O_8$.[14,15] A further complication arises because V ions in Sr-V-O with a 1:1 Sr:V ratio, including $Sr_2V_2O_7$, can adopt a 5+ oxidation state and can be thermodynamically favored in oxidizing environments.[16] The competing phases interrupt the epitaxy of $SrVO_3$ and are electrically insulating.[15,17] Minimizing the formation of competing phases is thus a key challenge for the epitaxial growth of $SrVO_3$.

Single-step approaches to the epitaxial growth of $SrVO_3$ approaches such as hybrid molecular beam epitaxy (hMBE), pulsed laser deposition (PLD), and radio-frequency (RF) sputtering require a specific set of gas pressures and substrate temperatures to synthesize phase-pure $SrVO_3$.[7,8,12,14,18-20] These methods employ elevated substrate temperature, typically 600 to 700 °C, to promote surface diffusion, posing a constraint for the design of deposition processes.

Here we report methods based on SPE that separate the kinetic phenomena of the deposition of $SrVO_3$, the control of the V oxidation state, and crystallization into different steps. The Sr:V ratio in particular can be readily controlled in SPE because the rate of evaporation of Sr and V compounds is low during the room-temperature deposition of the amorphous layer. Solid phase epitaxy also holds the prospect of crystallization over large-area amorphous substrates via nucleation at spatially isolated seed crystals, expanding the range of substrates from single crystals to include exfoliated nanosheets and other methods employing dispersed nucleation sites on non-templating substrates.[21,22] This approach combines favorable thermodynamic conditions and the distinct crystallization kinetics of SPE with scalable processes and opens new routes into the fabrication of multivalent complex oxides in non-planar forms.

The steps involved in the synthesis of epitaxial $SrVO_3$ on (001)-oriented $SrTiO_3$ using SPE are illustrated in Fig. 1(a). The process has two steps: the deposition of amorphous $SrVO_3$ by RF magnetron sputtering onto the room-temperature substrate and crystallization during heating in a reducing atmosphere. A crucial issue is that crystalline $SrVO_3$ nucleates only at the amorphous $SrVO_3$/crystalline $SrTiO_3$ interface. The $SrTiO_3$ substrates were prepared before depositing the $SrVO_3$ film using a previously described surface treatment.[23] $SrVO_3$ films were deposited from a stoichiometric $SrO$-$VO_2$ target (AJA International, Inc.) using an RF power of 30W, resulting in a deposition rate of 12.5 nm/hr. Films for which the structural, chemical, optical, and electrical transport properties are discussed below were deposited at a pressures ranging from 7.5-30 mTorr in a 5% $H_2$/95% Ar mixture and crystallized at 750 °C for 3 h in a 5% $H_2$/95% Ar atmosphere at ambient pressure and had thicknesses of 16 and 60 nm. The crystallization kinetics were investigated in ultra-high vacuum (UHV) using 30-nm-thick amorphous films deposited in a 25% $H_2$/75% Ar at a pressure of 18 mTorr. For all samples used in the preparation of this manuscript, the UHV and reductive gas environments served the same purpose of mitigating the formation of competing oxygen-rich phases by reducing the oxygen partial pressure.

Deposition of amorphous Sr-V-O layers was conducted in deposition vacuum system with base pressure of $10^{-7}$ Torr. Deposition after backfilling with Ar gas thus resulted in an oxygen partial pressure ($P_{O2}$) higher than the stability threshold for $SrVO_3$ and produced oxygen-rich amorphous Sr-V-O layers. A signature of the $V^{5+}$ oxidation state was observed in an x-ray photoelectron spectroscopy (XPS) study of the amorphous $SrVO_3$ layers, as described below in Section II. $H_2$ was introduced during the deposition with the intention of lowering the effective $P_{O2}$ in the deposition atmosphere. The addition of H2 during sputtering did not result in a significant variation in the $V^{5+}$ signal observed using XPS. Figure 1(b) shows chemical processes through which O is exchanged between the film and the gas atmosphere during crystallization, selecting the $V^{4+}$ oxidation state. Crystallization in a $H_2$-rich gas environment promotes the reduction of V via the formation of water from $H_2$ gas and O in the amorphous layer. The reduction of the surface produces a concentration gradient that drives the diffusion of excess O towards the surface. The reduction of the amorphous layers can also be accomplished in UHV, as described in the kinetics study below.

The figure of merit (FOM) $\Phi_{TC}$ for TCOs is $\Phi_{TC} = \frac{T_{opt}^{10}}{R_s}$.[24] Here $T_{opt}$ is the optical transmission at a wavelength of 550 nm, $R_s$ is the sheet resistance, and the exponent 10 arises from the arbitrary selection of 90% optical transmission.[24] Higher values of $\Phi_{TC}$ correspond to higher optical transmission and reduced sheet resistance and are thus desirable for TCO applications. As described in more detail below, $SrVO_3$ films crystallized using SPE with thicknesses of 16 nm and 60 nm have figures of merit $\Phi_{TC} = 5.9 \times 10^{-3}$ $\Omega^{-1}$ and $\Phi_{TC} = 3.8 \times 10^{-4}$ $\Omega^{-1}$ at 300 K, respectively. The values of $\Phi_{TC}$ for $SrVO_3$ films of similar thicknesses reported in the literature span three orders of magnitude, as summarized in Fig. 1(c).[8,14] The highest previously reported FOM for 60 nm SrVO$_3$ films is 5.5 × 10$^{-4}$ Ω$^{-1}$, based on reported values for films with similar thickness grown using PLD.[12] The FOMs reported for SPE are similar those for methods producing the lowest concentrations of structural defects, e.g. layers produced by PLD and hMBE, in part because the room-temperature resistivity is dominated by phonon scattering rather than by scattering at defects.

## II. RESULTS AND DISCUSSION

**A. Structural and Chemical Characterization.** The orientation and crystal structure of the crystallized SrVO$_3$ films were probed using x-ray diffraction. Figures 2(a) and (b) show the diffracted intensity along the [00L] direction of a 60 nm-thick crystallized SrVO$_3$ film collected using Cu K$_{\alpha 1}$ radiation. Figure 2(a) exhibits only 00L reflections of SrVO$_3$ and SrTiO$_3$, indicating that the SrVO$_3$ film is oriented with the [001] direction parallel to the substrate surface normal. Figure 2(b) shows a narrow range near the SrTiO$_3$ and SrVO$_3$ 002 reflections. The SrVO$_3$ 002 reflection is centered at $Q_z$ = 3.286 Å$^{-1}$, corresponding to an out-of-plane lattice parameter of 3.824 Å. The out of-plane direction thus exhibits a compressive strain of 0.5% in comparison with the unstrained SrVO$_3$ bulk lattice parameter, a$_{SrVO3}$ = 3.842 Å.[7] A coherently strained SrVO$_3$ layer on SrTiO$_3$ would have an out-of-plane lattice parameter 3.793 Å, given by $a_{SrVO3} = \frac{2\nu_{SrVO3}}{1-\nu_{SrVO3}}(a_{SrTiO3} - a_{SrVO3})$, where a$_{SrTiO3}$=3.905 Å is the lattice parameter of SrTiO$_3$ and $\nu_{SrVO3}$=0.28 is the previously observed Poisson ratio of SrVO$_3$ thin films.[7] The experimentally observed out-of-plane lattice parameter of SrVO$_3$ has a value between the bulk and coherently strained lattice parameters, indicating that the film is relaxed through the formation of structural defects during SPE. A reciprocal space map in the region spanning the 113 reflections of SrTiO$_3$ and SrVO$_3$ is shown in Fig. 2(c). The SrVO$_3$ 113 reflection is centered at $Q_z$ = 5.432 Å$^{-1}$ and $Q_{xy}$ = 2.303 Å$^{-1}$. With the assumption that the epitaxial strain yields a tetragonal distortion of the SrVO$_3$, the in-plane lattice parameter of the SrVO$_3$ layer is 3.894 Å. The in-plane lattice parameter is thus also consistent with a partial relaxation of the coherent strain in the SrVO$_3$ layer. The mosaic width of the SrVO$_3$ 002 reflection was 0.42° full-width-half-maximum (FWHM) which is consistent with the formation of defects through the relaxation of the epitaxial strain. Broadening due to mosaicity is also expected to contribute to the FWHM of the SrVO$_3$ 002 reflection.

The oxidation state of V in the amorphous and crystallized SrVO$_3$ films was analyzed using x-ray photoelectron spectroscopy (XPS) with Al K$_\alpha$ radiation. Figure 2(d) shows V 2p and O 1s binding energy spectra collected from amorphous and crystallized SrVO$_3$ films. The amorphous film exhibits V 2p peaks with prominent maxima centered at 517.0 eV and 525.3 eV, corresponding to the 2p$_{3/2}$ and 2p$_{1/2}$ transitions of V$^{5+}$.[25] The XPS spectrum of the amorphous film is described in more detail below. The V 2p$_{3/2}$ and 2p$_{1/2}$ peaks are at 515.4 eV and 523.6 eV, respectively, in the crystallized film, consistent with the V$^{4+}$ state. The multiple contributions to the V 2p peaks in crystalline SrVO$_3$ may indicate that multiple valences of V are present in the probed region but could also have a contribution due to many-body screening in metallic SrVO$_3$.[26] The spectra from both films were collected after removing 4 nm from the surface of the film using Ar$^+$ sputtering to remove oxidized species formed after transfer through air.

Scanning transmission electron microscope high-angle annular dark-field (STEM-HAADF) imaging was used to probe the structure of the SrVO$_3$/SrTiO$_3$ interface. Figure 3(a) shows a STEM-HAADF image collected from a 60-nm-thick crystallized SrVO$_3$ film. The image is shown in order to illustrate the alignment of the SrVO$_3$ crystal planes with corresponding planes in the SrTiO$_3$ substrate. Specifically, the planes containing the A site and B site atoms of the perovskite structure are continuous across the interface. Dislocations are expected to be present in 60 nm SrVO$_3$ films to account for the measured lattice parameters of the SVO that indicate it is not psuedomorphically strained. One dislocation is apparent in Fig. 3(a) with line direction $\xi = \langle 100 \rangle$. A Burgers circuit around the observable component of the dislocation is indicated by the blue-dashed box in Fig. 3(a) with the red segment indicating the Burgers vector $\mathbf{b} = a_\parallel [010]$, where $a_\parallel$ is the in-plane lattice parameter of $SrVO_3$.

The location of $SrVO_3$/$SrTiO_3$ interface was determined using the STEM-HAADF image intensity and chemical contrast from B site atoms across the interface. Figure 3(b) shows the average image intensity along paths containing B site atoms within the region contained in the yellow-dashed box in Fig. 3(a). Energy dispersive x-ray fluorescence spectroscopy (EDS) analysis using the Ti K and V K emission lines confirmed, albeit with lower spatial resolution, the location of the interface, as shown in Fig. 3(c). The EDS profiles in Fig. 3(c) are fit with error functions with an average FWHM of 1.2 nm.

**B. Electronic and Optical Properties.** The optical transmission and electrical resistivity were measured to evaluate the transparent conductor properties of the $SrVO_3$ films. The resistivity was measured with a four-probe van der Pauw geometry. The resistivity of 60-nm-thick $SrVO_3$ ranges from $6.6 \times 10^{-6}$ $\Omega$ cm at 5 K to $2.5 \times 10^{-5}$ $\Omega$ cm at room temperature and from $2.7 \times 10^{-5}$ $\Omega$ cm at 5 K to $5.2 \times 10^{-5}$ $\Omega$ cm at room temperature for 16-nm-thick films, as shown in Fig. 4(a). The residual resistivity ratio (RRR) $\rho_{300 K}/\rho_{5 K}$ was 2.0 for the 16-nm-thick layer and 3.8 for the 60-nm-thick layer. Resistivity measurements of the as-deposited amorphous layers were attempted, however, the resistivity of amorphous Sr-V-O was beyond the dynamic range of the instrument, indicating that the amorphous layers were highly resistive. The resistivity of the $SrVO_3$ layers produced by SPE at room temperature is similar to the most highly conductive $SrVO_3$ films produced by other techniques. The RRRs for $SrVO_3$ formed by SPE, however, are relatively low, only slightly higher than the value of 1.7 observed for epitaxial $SrVO_3$ grown via PLD, but much lower than for films grown by hMBE, which have RRR values up to 222.[7,18] The difference in the values of RRR among these synthesis methods likely arises from the sensitivity of the low-temperature resistivity to electron scattering at structural defects and impurities. The RRR values in $SrVO_3$ crystallized by SPE is consistent with the relatively high mosaic width of these films and the higher defect concentration in SPE layers in comparison with hMBE films.

The optical transmission of amorphous and crystallized $SrVO_3$ was evaluated using optical spectroscopy. Figure 4(b) shows a transmission spectrum for films of amorphous and crystallized $SrVO_3$ with thicknesses of 16 and 60 nm. The optical spectroscopy measurements compared the transmittance of a sample consisting of $SrVO_3$ on $SrTiO_3$ with that of an $SrTiO_3$ substrate. The maximum transmission for crystalline 16 and 60 nm $SrVO_3$ is 0.89 and 0.52, respectively, at a wavelength of 550 nm. An amorphous $SrVO_3$ layer 60-nm-thick had a higher transmission, 0.72, at the same wavelength. Under the assumptions that (i) the reflectances of the $SrVO_3$ and $SrTiO_3$ surfaces are identical and (ii) the reflectance of the $SrVO_3/SrTiO_3$ interface is zero, the optical absorption coefficient is $\alpha = -\frac{\ln(T_{opt})}{d}$. Here $T_{opt}$ is the optical transmittance of the $SrVO_3$ and $d$ is the thickness of the $SrVO_3$ layer. At 550 nm, the absorption coefficients for amorphous and crystalline $SrVO_3$ are $3.95 \times 10^4$ cm$^{-1}$ and $1.09 \times 10^5$ cm$^{-1}$, respectively. The value of the absorption coefficient in $SrVO_3$ layers crystallized by SPE is similar to values reported in dielectric spectra for $SrVO_3$ films grown with hMBE, $8.4 \times 10^5$ cm$^{-1}$.[8] Absorption coefficient spectra for crystalline and amorphous $SrVO_3$ are shown in Fig. 4(c).

**C. Properties of amorphous $SrVO_3$ and surface morphology of crystallized $SrVO_3$.** Figure 5(a) shows the grazing-incidence x-ray scattering pattern of an amorphous $SrVO_3$ film with a thickness of 60 nm, measured with an x-ray wavelength of 1.54 Å. The highest x-ray scattering intensity from amorphous $SrVO_3$ occurs at $2\theta_{peak} = 28.87°$, corresponding to $Q_{peak}=2.03$ Å$^{-1}$. The bond length corresponding to this intensity maximum, given by $2\pi/Q_{peak}$,[27] is 3.1 Å, corresponding to the separation of Sr and V in the SrVO$_3$ perovskite structure.

The density, roughness, and film thickness of amorphous SrVO$_3$ were measured with x-ray reflectivity using Cu k$_{\alpha 1}$ radiation. Figure 5(b) shows x-ray reflectivity curves acquired from 60-nm-thick samples of amorphous and crystallized SrVO$_3$. The density obtained from fits to the x-ray reflectivity curve of amorphous SrVO$_3$ was 4.33 g cm$^{-3}$. The root-mean-square (rms) surface roughness of the amorphous and crystallized SrVO$_3$ determined from the x-ray reflectivity measurements were 0.54 nm and 4.8 nm, respectively. The surface roughness and morphology of amorphous and crystallized SrVO$_3$ films were also probed using atomic force microscopy (AFM). AFM images of the surface morphology of amorphous and crystallized SrVO$_3$, are shown in Figures 4(c) and (d), respectively. The amorphous and crystalline SrVO$_3$ layers exhibit rms roughness values of 0.3 nm and 4.2 nm in AFM measurements, respectively.

**D. Activation Energy for SPE Crystallization.** The kinetics of the crystallization of SrVO$_3$ were probed using reflection high energy electron diffraction (RHEED). These experiments were conducted at a pressure of $1\times10^{-9}$ Torr using amorphous SrVO$_3$ films with an initial thickness of 30 nm, selected for the RHEED to reduce the time required for crystallization. The vacuum environment of the RHEED study is highly reducing and results in the reduction of the V oxidation state during crystallization. The RHEED pattern obtained from amorphous SrVO$_3$ exhibits diffuse scattering with no crystalline reflections. At 600°C, reflections appear in the RHEED patterns after an elapsed time, during which amorphous/crystalline interface propagates from the substrate/film interface to the surface. RHEED patterns acquired with the incident beam along <110> and <100> exhibit transmission reflection in the SrVO$_3$ {100} or {110} family, respectively, as shown in Figs. 5(a) and (b).[28] The lattice spacings derived from Figs. 5(a) and (b) correspond to an in-plane lattice parameter of 3.8 ± 0.3 Å, matching SrVO$_3$. The RHEED patterns of the crystallized SrVO$_3$ indicate that the surface is rough enough to yield transmission diffraction, consistent with the x-ray reflectivity and AFM results. The roughness of the SrVO$_3$ films produced by SPE in this case is higher than for SrVO$_3$ films grown by conventional direct epitaxy (*e.g.* hMBE or PLD), resulting in spotty, transmission-like RHEED patterns characteristic of a rough surface.[7,28]

Electrons contributing to the RHEED diffraction pattern are scattered from a near-surface region with a thickness of approximately 1 nm, which allows the RHEED experiments to be used to determine the time at which the amorphous/crystalline interface reaches the surface, as illustrated in Fig. 6(c). We define $t$ to be the time at which the $010$ and $0\bar{1}0$ RHEED reflections appear. The crystallization velocity is given by $v=d/t$. Experimental uncertainty arises from the rate of temperature increase during heating, which ranges from 5 to 7 °C/s, and from the absolute accuracy of the temperature measurement, approximately 25 °C. The values of $v$ range from 0.004 to 0.2 nm/s in from 625 °C to 735 °C, as shown in Fig. 6(d).

The activation energy extracted from a fit of an Arrhenius temperature dependence to Fig. 6(d) is 2.7 eV. The activation energy inferred from Fig. 6(d) for SrVO$_3$ is significantly higher than previous observations of the kinetics of the crystallization of SrTiO$_3$ in air in SPE and in lateral epitaxial crystallization in a similar temperature regime.[21,23,29] The activation energy for the crystallization of SrTiO$_3$, however, depends sensitively on the gas atmosphere, with significantly higher activation energies in the absence of water vapor and the associated introduction of H at the amorphous/crystalline interface. The 2.7 eV activation energy for SrVO$_3$ crystallization is consistent with the activation energy of 2.1 eV observed for the crystallization of SrTiO$_3$ in vacuum.[29] It is also possible that the H-induced increases in crystallization rates would also be observed for SrVO$_3$. In that case, the selection of the gas atmosphere would provide separate control of the V valence via the O activity, and crystallization rate via the H concentration.

**E. Thermodynamic stability of SrVO₃.** The thermodynamics of the formation of SrVO$_3$ were studied using density functional theory (DFT) total energy calculation and subsequent analysis of multicomponent phase stability. The DFT total energy of SrVO$_3$ was obtained using the GGA+$U$ methods.[30] The multicomponent phase stability calculations were performed using the total energy calculated using DFT and the analysis package pymatgen.[31] The phase stability calculations were performed for the Sr-V-O system open to reaction with O$_2$ gas, with a temperature- and oxygen pressure-dependent oxygen chemical potential.[30]

The stability of the SrVO$_3$ phase is represented in Fig. 7 as the convex hull energy, the energy difference between SrVO$_3$ (the reactant) and the linear combination of most stable phases constituting the surface of the phase diagram at that pressure and temperature (products). A convex hull energy of zero in Fig. 7 would indicate that SrVO$_3$ is formally stable and resides on the phase diagram at a specific combination of $T$ and $P$. A positive convex hull energy indicates that SrVO$_3$ is formally unstable but may still be obtainable in practice, with the final product formation depending on the kinetics of formation of competing phases. Figure 7 indicates that SrVO$_3$ is most stable at high temperatures ranging from about 825-1525 °C and low oxygen pressures ranging from 10$^{-20}$ Torr (at T≈825 °C) to 10$^{-8}$ Torr (at T≈1525 °C). In the range of experimental conditions considered in the H$_2$/Ar crystallization and kinetic studies using RHEED, the oxygen partial pressure is on the order of 10$^{-9}$ Torr or lower. Under these experimental conditions the energy of SrVO$_3$ above the convex hull is on the order of 100 meV atom$^{-1}$, indicating that, under these conditions, SrVO$_3$ may be classified as metastable and may exhibit long-term kinetic stability. Under more oxidizing conditions, i.e. lower $T$ and higher oxygen partial pressure, the upper left of Fig. 7, the SrVO$_3$ convex hull energy increases and V$^{5+}$ is favored instead of the desired V$^{4+}$.

Depending on the precise processing scheme adopted, these oxidizing conditions are expected to yield either $Sr_2V_2O_7$ or $Sr_3V_2O_8$ secondary phases. Processing conditions at relatively high temperatures favor decomposition into binary oxides.

## III. CONCLUSION

Solid-phase crystallization of amorphous $SrVO_3$ can be used to form epitaxial films of $SrVO_3$ on $SrTiO_3$ substrates. The stabilization of the cubic phase during crystallization requires a controlled gas environment selected to favor the $V^{4+}$ oxidation state and promote the formation of the cubic perovskite phase by suppressing the formation of competing insulating phases such as $Sr_2V_2O_7$ and $Sr_3V_2O_8$. Thin $SrVO_3$ films created with SPE have transparent conductor figures of merit of similar magnitude to optimized films produced with PLD and hMBE grown $SrVO_3$ because the TCO properties are somewhat insensitive to the slightly higher concentration of structural defects resulting from SPE.

The optimum thickness for $SrVO_3$ films grown by SPE will be less than the 16 and 60 nm thickness at which the basic characterization was performed here, in order to increase the optical transmission. With the assumption that the absorption coefficient and resistivity are independent of the sample thickness, the thickness with the maximum $\Phi_{TC}$ is $1/10\alpha$.[24] With $\alpha = 1.09 \times 10^5$ cm$^{-1}$ from above, the optimum thickness of SPE-derived $SrVO_3$ TCOs is 9.5 nm. An $SrVO_3$ film with that thickness would have $\Phi_{TC} = 2.4 \times 10^{-2}$, approximately equal to the optimum value for $SrVO_3$ films produced by hMBE, which have a similarly thin optimum thickness. The reduction in thickness to 9.5 nm from the thicknesses of $SrVO_3$ films described here may, however, induce significant surface and interface scattering, even at room temperature and the actual $\Phi_{TC}$ at this thickness, and the overall optimum thickness, may not meet the simple predictions.

Beyond the crystallization of $SrVO_3$ on single-crystal $SrTiO_3$ substrates, the use of SPE to produce high-figure-of-merit transparent conductors raises the possibility that TCO films can be produced with large areas and in geometries for which vapor-phase epitaxial growth is not possible. The kinetic phenomena and crystallization environment of $SrVO_3$ occur in conditions that are compatible with the introduction of seed crystals, the pre-patterning of three-dimensional structures, and the production of large areas of crystalline TCO films over amorphous substrates. The water solubility of $SrVO_3$ will need to be considered in the layer formation process and fabrication and application of devices that incorporate crystalline $SrVO_3$.[32] Patterning for oxide substrates can require exposure to water or other wet chemical processing steps that may degrade crystalline $SrVO_3$ layers. Therefore, deposition of amorphous $SrVO_3$ onto a pre-patterned substrate that sets the orientation and direction of the crystallization of the amorphous layer into a final geometry, or a similar dry fabrication process may be necessary to preserve the structural and chemical state of $SrVO_3$. More broadly, these results demonstrate a potential route toward low temperature crystallization of epitaxial TMO thin films and are an important step toward the growth of large surface area films of transparent conducting oxides.

**Acknowledgements**

This research was primarily supported by NSF through the University of Wisconsin Materials Research Science and Engineering Center (DMR-1720415). Support for L. Lin and R. Jacobs was provided by the Defense Advanced Research Projects Agency (DARPA) through the Innovative Vacuum Electronic Science and Technology (INVEST) program.

**Conflict of Interest**

The authors declare no conflict of interest.

**Fig. 1**. (a) Solid phase epitaxial growth of SrVO$_3$. (i) Amorphous SrVO$_3$ deposited at room temperature on SrTiO$_3$ (001). (ii) Crystallized SrVO$_3$ after heating and epitaxial relationship between SrVO$_3$ and SrTiO$_3$. (b) Oxygen transport and exchange through the free surface crystallization in a reductive gas atmosphere. (c) Transparent conductor figures of merit for 16 and 60 nm-thick SrVO$_3$ synthesized by SPE (this work), and reported figures of merit for SrVO$_3$ films grown by hMBE (diamonds),[8] PLD (upward,[4] forward,[20] and downward[12] pointing triangles), and sputtering (circles).[14]

**Fig. 2**. (a) θ-2θ diffraction pattern showing 00L x-ray reflections for a crystallized epitaxial SrVO$_3$ film and the SrTiO$_3$ substrate, acquired at x-ray wavelength 1.5406 Å. (b) Diffraction profile near the 002 reflections. (c) Reciprocal space map in the region of the SrTiO$_3$ and SrVO$_3$ 113 x-ray reflections. $Q_z$ and $Q_{xy}$ represent the components of the x-ray wavevector along the surface normal and [110] in-plane directions, respectively. (d) XPS core level spectra for V 2p and O 1s in the amorphous precursor film (red) and a crystallized SrVO$_3$ film (blue). The spectrum for the crystallized layer is shifted vertically by 0.1.

**Fig. 3.** (a) STEM-HAADF image of the 60 nm film crystallized at 750 °C for 3 h. (b) Average image intensity along the B site columns ±3 nm from the interface. (c) EDS intensity profile for Ti and V measured normal to the interface at the location of the black arrow in (a).

**Fig. 4.** (a) Electrical resistivity of crystallized 16 nm (open circles) and 60 nm (open squares) SrVO$_3$ films. (b) Optical transmission spectrum 60-nm-thick amorphous (open triangles) and crystallized 16 nm (open circles) and 60 nm (open squares) SrVO$_3$ films. (c) Optical absorption coefficient spectra for amorphous and crystallized SrVO$_3$ films calculated from the transmission spectra in (b) under assumptions described in the text.

**Fig. 5.** (a) X-ray scattering intensity from amorphous SrVO$_3$, collected with x-ray wavelength 1.54 Å. (b) X-ray reflectivity curves for amorphous (red) and crystallized (blue) SrVO$_3$, collected with an x-ray wavelength of 1.5406 Å. X-ray reflectivity models with parameters given in the text are shown as solid lines. The reflectivity of the crystallized layer has been multiplied by 0.1. (c) AFM height map of amorphous SrVO$_3$. (d) AFM height map of crystallized SrVO$_3$.

**Fig. 6.** RHEED patterns acquired after crystallization at 735 °C with the incident beam oriented along (a) <110> and (b) <010> directions. Integrated intensity profiles are shown below each pattern. (c) Schematic definition of crystallization time *t*, at which the crystallization interface reaches the ~1 nm depth probed by the RHEED experiment. (c) Temperature dependence of the crystallization velocity with a fitted activation energy of 2.74 eV.

**Fig. 7.** Convex hull energy for the Sr-V-O system as a function of temperature and oxygen partial pressure calculated using DFT. Higher convex hull energy values indicate a stronger driving force for reaction of SrVO$_3$ to energetically favored competing phases.

**Figure 1**

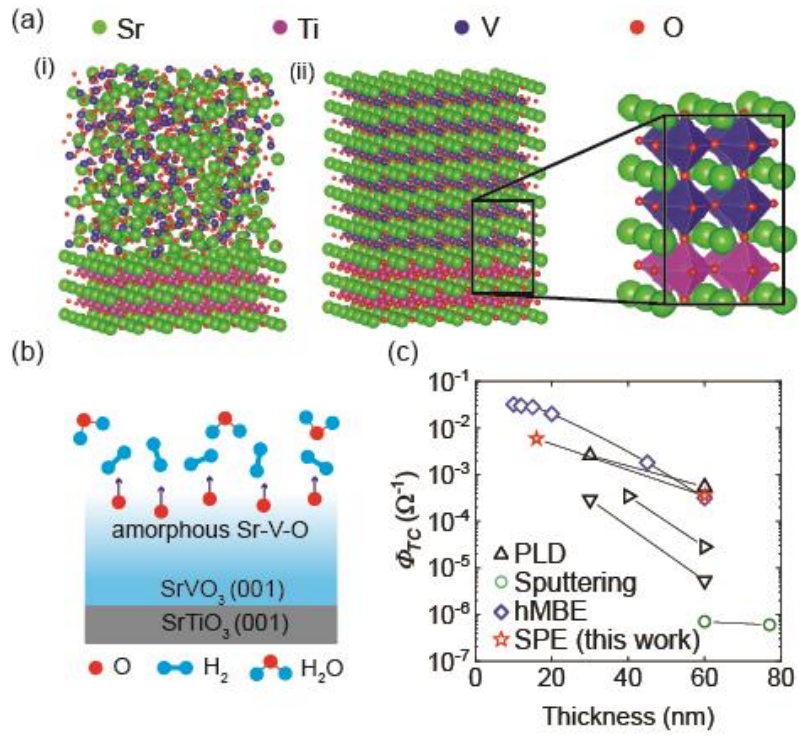

**Figure 2**

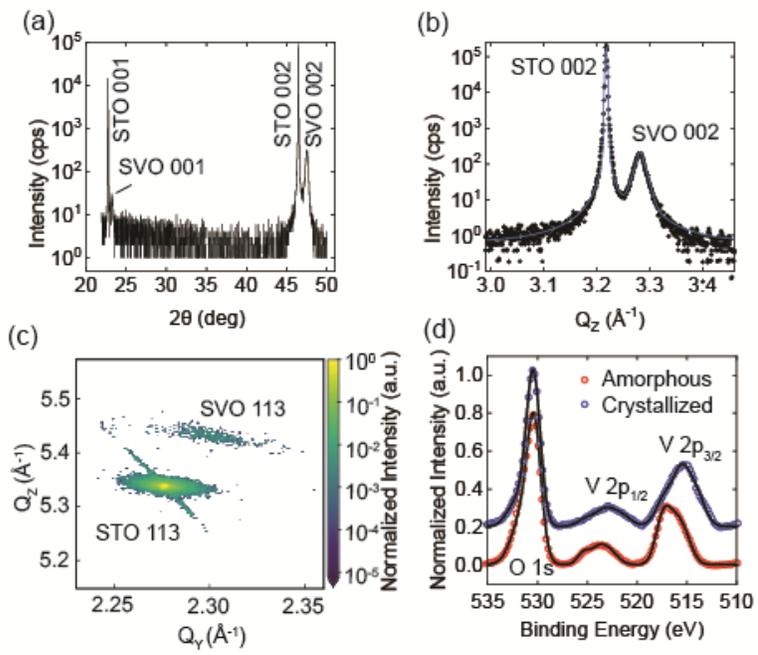

**Figure 3**

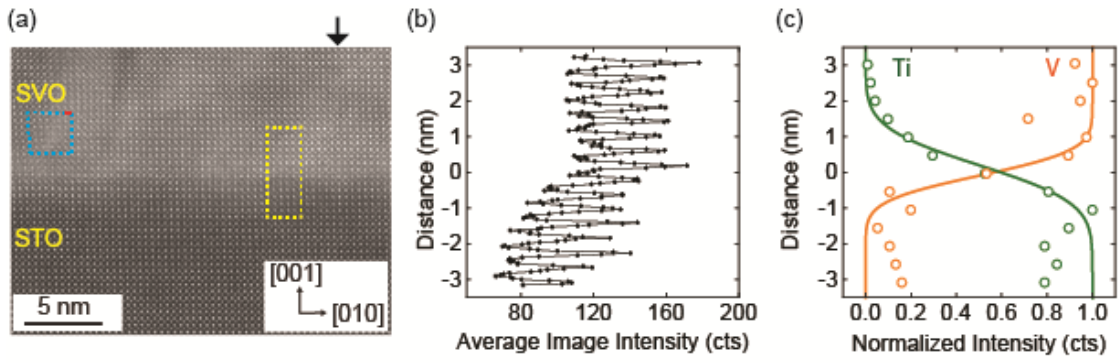

**Figure 4**

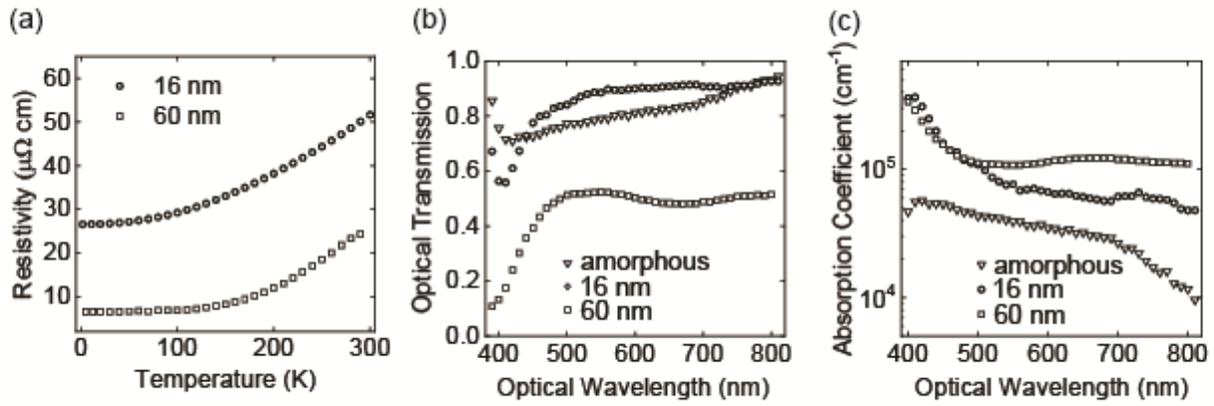

**Figure 5**

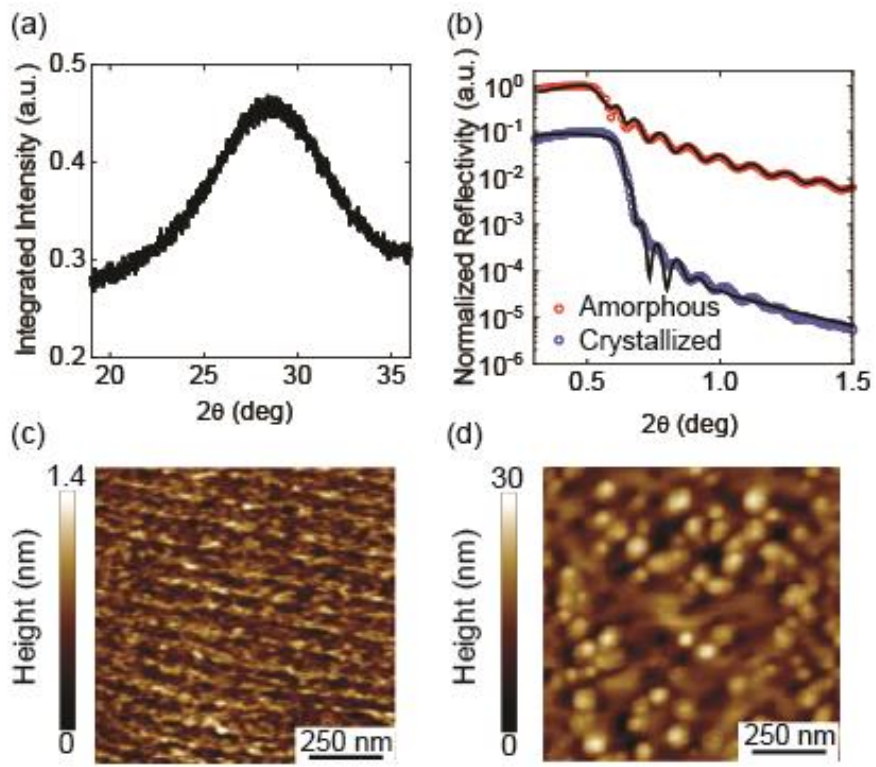

**Figure 6**

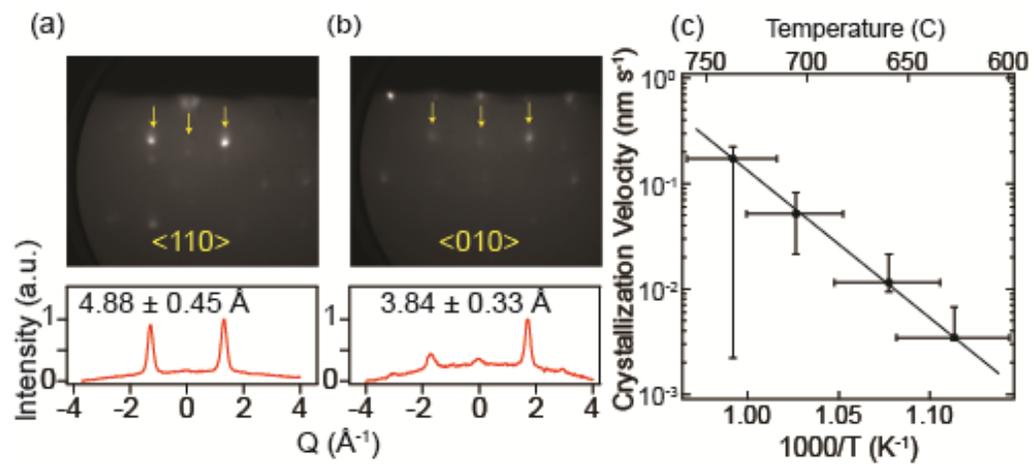

**Figure 7**

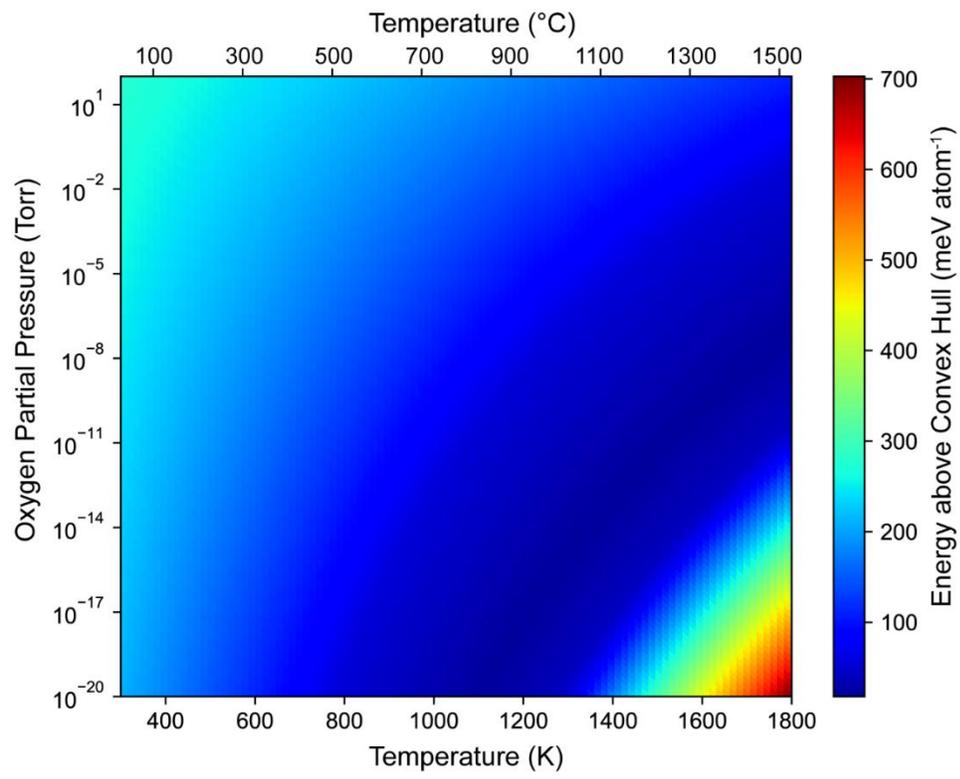